\documentclass{UApreprint}
\usepackage{graphics}
\usepackage{fullpage}
\usepackage{verbatim}
\usepackage[tight]{subfigure}
\usepackage{amssymb}
\usepackage{citesort}
\newcommand{\be}{\begin{equation}}
\newcommand{\ee}{\end{equation}}
\newcommand{\bea}{\begin{eqnarray}}
\newcommand{\eea}{\end{eqnarray}}
\newcommand{\bean}{\begin{eqnarray*}}
\newcommand{\eean}{\end{eqnarray*}}
\newcommand{\bi}{\begin{itemize}}
\newcommand{\ei}{\end{itemize}}
\newcommand{\bdm}{\begin{displaymath}}
\newcommand{\edm}{\end{displaymath}}

\newcommand{\lag}{{\mathcal L}}

\newcommand{\meg}{{\mu\to e\gamma}}

\newcommand{\Ye}{\mathbf{Y}_e}
\newcommand{\Yn}{\mathbf{Y}_\nu}
\newcommand{\Yd}{\mathbf{Y}_d}
\newcommand{\Yu}{\mathbf{Y}_u}

\begin{document}
%\draft

\title{Triviality and the (Supersymmetric) See-Saw}
\author{Bruce A. Campbell$^1$ and David W. Maybury$^2$}
\address{$^1$Department of Physics, Carleton University, 
1125 Colonel By Drive, \\ Ottawa ON K1S 5B6, Canada \\
$^2$Rudolf Peierls Centre for Theoretical Physics, 
University of Oxford, 1 Keble Road, Oxford OX1 3NP, United Kingdom}

\date{}

\abstract{For the $D=5$ Majorana neutrino mass operator to have a see-saw
ultraviolet completion that is viable up to the Planck scale, the see-saw
scale is bounded above due to triviality limits on the see-saw couplings. 
For supersymmetric see-saw models, with realistic neutrino mass textures, 
we compare constraints on the see-saw scale from triviality bounds, with
those arising from experimental limits on induced charged-lepton flavour 
violation, for both the CMSSM and for models with split supersymmetry.}

\archive{}
\preprintone{OUTP-06 06P}
\preprinttwo{}

\submit{}

\maketitle

%%%%%%%%%%%%%%%%%%%%%%%%%%%%%%%%%%%%%%%%%%%%%%%%%%%%
%%%%%%%%%%%%%%%%%%%%%%%%%%%%%%%%%%%%%%%%%%%%%%%%%%%%

\section{Introduction}

The standard model of elementary particles and their interactions has provided a singularly successful description of nature, as it presents itself in the laboratory. As well as providing the underlying theory of the physics of elementary particles, as compiled by the Particle Data Group \cite{Eidelman:2004wy}, it forms the reductive foundation of the other fields of physics built upon it, and thus the other fields of science that in turn depend on physics for their basic inputs. However, it is widely expected that the standard model does not offer the complete story; it appears ad hoc, and requires a choice of gauge groups and matter representations, as well as the input of nineteen carefully tuned constants of various kinds. More importantly, careful experimental efforts have revealed new physical phenomena in the form of cosmological dark energy and matter, and the flavour oscillations of massive neutrinos, which lack a description by the standard model.

If we consider the incorporation of gravitation in the standard model using Einstein's theory of gravity as an effective field theory, valid below the Planck scale ($M_P \approx 2\times 10^{18}\hspace{1mm}\mathrm{GeV}$), then dark energy can be described in terms of the lowest dimensional operator which one can covariantly include, a cosmological constant term. The puzzle this term presents is not it's inclusion (it will be generated in the renormalization group flow down from the Planck scale), but rather why the cosmological constant term appears 120 orders of magnitude smaller than the dimensional analysis of the flow of this relevant operator would suggest. On the other hand, the description of cold dark matter (assuming that it is particle dark matter as opposed to, say, small black holes), appears to require extension of the standard model field content. As our present knowledge of this component of nature is limited to its gravitational effects in cosmological and galactic dynamics, the required extension of the standard model particle content remains poorly constrained. However, achieving the correct dark matter density with thermal relics would strongly suggest that the new particles make up part of the effective field theory well below the Planck scale with particle masses close to the electroweak scale. 

The recent observations of neutrino flavour oscillations \cite{Cleveland:1998nv,Ahmad:2001an,Ahmad:2002jz,Ahmad:2002ka,Fukuda:1998mi,Fukuda:1998ah,Fukuda:2000np,Fukuda:2001nk,Eguchi:2002dm,Araki:2004mb,Ahn:2002up,Aliu:2004sq,Hampel:1998xg,Anselmann:1995ag,Abdurashitov:2002nt,Abdurashitov:1999bv,Abdurashitov:1994bc} provide compelling evidence for neutrino mass. In the absence of neutrino mass the standard model perturbatively conserves lepton flavor, since the charged lepton Yukawa matrix of the standard model with massless neutrinos can be diagonalized with respect to the gauge interactions. Considering the standard model as an effective field theory, the requisite neutrino mass terms can in general be introduced without extension of the standard model field content. From the low-energy point of view, neutrino mass terms appear through a non-renormalizable dimension-five operator constructed from standard model fields, 
%%%%%%%%%%%%%%%%%%%%%%%%%%%%%%%%%%%%%%%%%%%%%%%
\be 
\mathcal{O}_{\nu} = 
{\lambda_{ij}}L^iL^jHH + \mathrm{h.c.} 
\label{nu_op} 
\ee 
%%%%%%%%%%%%%%%%%%%%%%%%%%%%%%%%%%%%%%%%%%%%%%
where $i$ and $j$ refer to generation labels, $\lambda_{ij}$ define arbitrary coefficients with dimensions of inverse mass, and $SU(2)$ indices have been suppressed. Since there exists only one dimension-five operator that can be constructed from a gauge-invariant, Lorentz-invariant combination of standard model fields, and since dimension five gives the leading irrelevant operators, Wilsonian reasoning predicts that the first laboratory indications of physics beyond the renormalizable (relevant and marginal) standard model interactions should arise from effects of neutrino mass, as confirmed by experiment. The experimental lower bound $\Delta(m^{2}) > 0.0015 (\mathrm{eV}^{2})$ for the oscillation $\nu_{\mu} {\not\rightarrow} \nu_{\mu}$ implies that in a mass-diagonal basis for the neutrinos, and with $\langle H \rangle = 250/{\surd{2}}$, there exists a neutrino mass eigenstate for which (diagonal) $\lambda > 1.6 \times 10^{-15} (\mathrm{GeV}^{-1})$.

We have added the irrelevant operator $\mathcal{O}_{\nu}$ to the standard model in order to describe the low-energy neutrino Majorana mass terms. Because $\mathcal{O}_{\nu}$ is an (infrared) irrelevant operator (ie. nonrenormalizable), its effects in physical processes grow as a power of the energy. Unitarity constraints on scattering cross sections at high energies will then impose a limit on the range of validity of this effective description of the physics responsible for neutrino mass and oscillation, and give an upper bound on the energy at which this effective description must be replaced by its UV completion. We use unitarity constraints on the cross-section for $\nu\hspace{1mm} H \rightarrow \nu^c \hspace{1mm} H$ for the largest neutrino mass eigenstate to place an upper bound on the scale at which the neutrino mass operator $\mathcal{O}_{\nu}$ must be replaced 
by the underlying UV completion from which it arises. As $\mathcal{O}_{\nu}$ is a local operator, the scattering process occurs in the s-wave, and we have a partial-wave unitarity constraint on the cross-section. Comparing this to the Born approximation for the cross-section we find that:
\be
\sigma = \frac{2}{\pi}\left({\lambda_{ij}}\right)^2 
\leq 
\frac{8\pi}{s}.
\ee
%%%%%%%%%%%%%%%%%%%%
For energies such that this comparison indicates Born estimates of the 
cross section that violate unitarity, we have a breakdown of perturbative 
calculability of the theory. In general, this is understood as an indication that one is leaving the range of validity of the effective theory in question, and that at these energy scales one must replace the effective theory by a more fundamental high energy theory, which acts as its ultraviolet completion (indeed, precisely this chain of reasoning led us to understand that at energies of order hundreds of GeV Fermi's theory of the weak interactions would need to be replaced by its ultraviolet completion, the Glashow-Salam-Weinberg theory).  

Since the large $\Delta(m^{2})$ from the atmospheric $\nu_{\mu} {\not\rightarrow} \nu_{\mu}$ oscillations established $\lambda > 1.6 \times 10^{-15} (\mathrm{GeV}^{-1})$ for the largest mass eigenstate, partial wave unitarity for scattering of that mass eigenstate requires that:
%%%%%%%%%%%%%%%%%%%%%%%%%%%%%
\be
\label{UV_complete}
\sqrt{s} \leq 3.8 \times 10^{15} \mathrm{GeV}. 
\ee
%%%%%%%%%%%%%%%%%%%%%%%%%%%
So, for the dimension-five neutrino-mass operator, $\mathcal{O}_{\nu}$, required by the present neutrino oscillation data, unitarity constraints require its incorporation in a UV completion of the theory at an energy scale no higher than that given by eq.(\ref{UV_complete}), which is well below the Planck scale at which the gravitational sector of the standard model requires UV completion. \footnote {The inverse of the present argument appears in \cite{Yan}. There the authors used the requirement of Planck-scale partial-wave unitarity to bound neutrino mass. Here we use the {\it observed} neutrino mass to determine when partial-wave unitarity requires that the neutrino mass operator $\mathcal{O}_{\nu}$ be replaced by its ultraviolet completion.} The relatively large neutrino mass squared differences obtained from the atmospheric oscillation results of (Super)Kamiokande, and the K2K experiment, unambiguously indicate a breakdown of the standard model description of neutrino mass physics at a sub-Planckian (but typically high) energy  scale. While eq.(\ref{UV_complete}) provides an upper bound on the scale of the UV completion, it does not specify a lower bound. Thus, the operator given in eq.(\ref{nu_op}) may in principal have an explanation anywhere from the weak scale up to the bound in eq.(\ref{UV_complete}).
%%%%%%%%%%%%%%%%%%  

The see-saw mechanism provides, perhaps, the most elegant UV completion that generates the operator $\mathcal{O}_{\nu}$. In this scenario, the standard model extends to include three massive gauge-singlet (right-handed) neutrino fields that interact with the left-handed lepton doublet through a Dirac-type Yukawa interaction. Integrating out the right-handed neutrinos at the see-saw scale generates the light neutrino masses, resulting in the operator $\mathcal{O}_{\nu}$. However, as we will discuss in the next section, by itself the see-saw mechanism does not necessarily provide a fully viable UV completion all the way to the Planck scale. The see-saw mechanism involves Yukawa interactions which are not asymptotically free. If the mass scale of the see-saw is sufficiently large, the initial size of these Yukawa couplings at the see-saw scale will be large enough that, under renormalization group evolution in the see-saw extended standard model, the Yukawa couplings will be driven to a Landau pole at energies below the Planck scale \cite{Casas:1999ac}. In section 2 we will analyze this effect, and show that requiring the see-saw mechanism not to suffer a Landau pole below the Planck scale will place an upper bound on the scale at which the see-saw is introduced, and that the 
corresponding bound is stronger than that following from partial-wave unitarity alone.  

These considerations become of particular interest in supersymmetric extensions of the standard model (SSM) where the soft supersymmetry-breaking is introduced at the Planck scale, such as models based on minimal supergravity (MSUGRA). In these models no new observable sector sub-Planckian fields are required to mediate the supersymmetry breaking, so the effective observable sector field content is just the SSM, and the soft supersymmetry breaking terms come as part of the observable sector effective field theory emerging from the Planck scale, and renormalize to low energies. Models in this class, if they are endowed with R-parity or an equivalent gauged discrete symmetry, have a stable lightest supersymmetric partner (LSP) which is often a neutralino of the kind necessary to have a thermal relic dark-matter density in the cosmologically required range \cite{Ellis:2003cw,Battaglia:2003ab,Ellis:2003si,Baer:2003yh,Baer:2003wx}. As such, they solve the cosmological cold dark matter problem which is posed by experiment but not answered by the standard model. So in the absence of neutrino masses, the supersymmetric extension of the standard model is a potentially viable effective description of observable sector physics up to the the Planck scale. 

In the presence of neutrino masses, our considerations above indicate that we need to introduce new fields, such as the see-saw singlets, to retain the possibility of a viable theory up to the Planck scale. But this implies an interval of RGE running of the soft masses below the Planck scale down to the see-saw scale, during which the see-saw Yukawa interactions will induce flavour violations in the soft breaking scalar mass-squared terms, which at low energy will induce charged-lepton flavour violating decays. Experimental limits on these decays will (depending on the size of the low-energy superpartner masses) give us upper bounds on the size of the Yukawa superpotential interactions involved. The Yukawa couplings in turn are responsible for the neutrino masses via the see-saw, so the observed neutrino masses will give us upper bounds on the see-saw scale. It is now interesting to determine whether we arrive at stronger bounds on the scale of the neutrino see-saw from the requirement that we avoid a Landau pole below the Planck scale (the ``triviality" bound), or from the requirement that the induced charged-lepton flavour violation not exceed present experimental limits. As we will see below, the strength of each bound depends on the nature and scale of the soft supersymmetry-breaking masses that are incorporated in particular supersymmetric models. In particular, we will examine the low-energy impact of susy see-saw (non)-triviality, on charged-lepton flavor violation in parameter ranges of the MSSM, and Split Supersymmetry \cite{Arkani-Hamed:2004fb,Giudice:2004tc,Arkani-Hamed:2004yi}, consistent with all laboratory and cosmological observations.        
 
%%%%%%%%%%%%%%%%%%%%%%%%%%%%%%%%%%%%%%%%%%%%%%%%%%%%%%%%%%%%%%%
%%%%%%%%%%%%%%%%%%%%%%%%%%%%%%%%%%%%%%%%%%%%%%%%%%%%%%%%%%%%%%%

\section{The See-saw Mechanism and Triviality}

%%%%%%%%%%%%%%%%%%%%%%%%%%%%%%%%%%%%%%%%%%%%%%%%%%%%%%%%%%%%%%%

\subsection{The Non-Supersymmetric Case}

As an illustrative example, consider the non-supersymmetric see-saw extension of the standard model. The lepton sector Lagrangian is extended by the terms:
\be
\lag= {\Yn}^{ij}\bar L_{j} N_i H^* + 
\frac{1}{2} \mathcal{M}^{ij} \overline{N^c}_i N_j + \mathrm{h.c.}
\ee
where $\Yn$ denotes Yukawa couplings and $\mathcal{M}$ denotes the singlet mass matrix (again, SU(2) indices are suppressed). Upon integrating out the $N_i$ fields at the see-saw scale, we obtain the light Majorana neutrino mass matrix after electroweak symmetry breaking,
\be
{\bf m_\nu}={\bf M_D}^T\mathcal{M}^{-1} {\bf M_D}
\label{nu_mass}
\ee
where ${\bf M_D}= \langle H^0 \rangle \Yn$. In general, ${\bf m_\nu}$ is not diagonal, leading to a mismatch between the weak and mass eigenstates. The basis of our argument is the lack of asymptotic freedom exhibited by renormalization group equations (RGEs) associated with the see-saw parameters,
\bea
\frac{d\Yn}{d t} &=& \frac{1}{16\pi^2} \Yn \left(\frac{3}{2} 
\Yn^\dagger \Yn + \mathrm{Tr}(\Yn^\dagger \Yn) -\frac{3}{2} 
\Ye^\dagger \Ye + \mathrm{Tr}(\Ye^\dagger \Ye)\right. \nonumber \\
&& + \left.3\mathrm{Tr}(\Yu^\dagger \Yu) + 3 \mathrm{Tr} 
(\Yd^\dagger \Yd) -\frac{9}{20}g_1^2 - \frac{9}{4} g_2^2\right).
\label{nsusy_rge}
\eea
where $t$ is the log of the renormalization scale. 

We may crudely illustrate the development of the Landau pole by assuming only one generation and ignoring all but the first two terms in eq.(\ref{nsusy_rge}), giving the approximate solution
\be
Y_\nu^2(t) \approx - \frac{Y_{\nu}(0)^2}{1 - Y_{\nu}(0)^2 \frac{5}{16\pi^2}t}
\label{nsusy_solt}
\ee
where $Y_{\nu}(0)$ is the initial Yukawa coupling at the see-saw scale. We see that for sufficiently large $t$ the one loop approximate solution of 
the RGE diverges (the Landau pole), and that for large initial $Y_{\nu}(0)$ the required $t$ becomes small. If the Landau pole is a real feature of the full non-perturbative renormalization group behaviour of the theory ({\it e.g.} as is rigorously known to be the case in $\lambda \phi^{4}$ theory in four dimensions), then to be valid as an effective theory up to a given energy scale, the low-energy couplings must be small enough not to produce the Landau pole by renormalization group evolution in the presumed range of validity of the theory. As one attempts to take the ``continuum" limit (effective theory valid to arbitrarily large energy), this bound on the low energy couplings approaches zero and the theory becomes trivial (free). Hence limits on couplings in effective theories suffering a Landau pole in the ultraviolet are known as ``triviality" bounds. It is precisely this reasoning on the quartic Higgs self-coupling $\lambda$ in the standard model, that bounds $\lambda$ from above (and hence the Higgs mass) by a ``triviality" bound that depends on the energy scale up to which the Higgs must provide an effective field theory description of the physics. In general, we expect that if the perturbative RGE evolution into the ultraviolet is afflicted with a Landau pole, that the pole signals a breakdown of the effective field theory within which we have been working, and indicates the necessity that it be replaced by its UV completion at an energy scale corresponding to the onset of non-perturbative behaviour. 

At present solar, reactor, and long base-line neutrino oscillation data
\cite{Cleveland:1998nv,Ahmad:2001an,Ahmad:2002jz,Ahmad:2002ka,Fukuda:1998mi,Fukuda:1998ah,Fukuda:2000np,Fukuda:2001nk,Eguchi:2002dm,Araki:2004mb,Ahn:2002up,Aliu:2004sq,Hampel:1998xg,Anselmann:1995ag,Abdurashitov:2002nt,Abdurashitov:1999bv,Abdurashitov:1994bc} indicate,
\bea
|\Delta m^2_{12}| = (2.6 - 16) \times 10^{-5} \hspace{1mm} 
\mathrm{eV}^2 \nonumber \\
|\Delta m^2_{23}| = (1.5 - 3.7)\times 10^{-3} \hspace{1mm} \mathrm{eV}^2.
\eea
Assuming a normal hierarchy, $m_3 \gg m_2 \gg m_1$, the data suggest the heaviest Majorana neutrino mass of $m_3 \approx .05$ eV. For definiteness, we take $m_3 = 5\times 10^{-2}$ eV, $m_2= 5\times 10^{-3}$ eV and $m_1\approx 0$ throughout. Using $m_3 = .05 \hspace{1mm} \mathrm{eV}$ and $\langle H^0 \rangle = 174 \hspace{1mm} \mathrm {GeV}$, we find that at the see-saw scale,
\be
\label{ratio}
\frac{|Y_\nu|^2}{M} = 1.6 \times 10^{-15} \hspace{1mm} 
{\mathrm{GeV}}^{-1}
\ee
(where for simplicity we have here assumed a single diagonal mass eigenvalue $M$ for the see-saw partner of the largest mass eigenvalue neutrino state; for the supersymmetric case discussed in the next subsection we will consider the generation structure in detail). In this letter we ignore radiative corrections on the induced dimension five operator eq.(\ref{nu_op}) below the see-saw scale. These effects can be significant \cite{Antusch:2001ck,Antusch:2003kp} when matching the observed neutrino oscillation data, however, these effects have only a subleading influence on our results and conclusions. While empirical data establish the ratio of the square of the Yukawa coupling to the scale $M$, they do not restrict the terms individually. However, we see that in order to hold the value of the right-hand side (which is determined by the largest neutrino mass eigenvalue) of 
eq.(\ref{ratio}) fixed, we need to scale up $|Y_\nu|^2$ as the first power of the see-saw scale ${M}$. If the scale of the neutrino singlet masses 
$M$ is too large, then the neutrino Yukawa coupling required by the  observed  neutrino masses will already be so large at the see-saw scale, 
that under its non-asymptotically-free running it will hit a Landau pole at an energy below the Planck scale. So the requirement that the standard model augmented by the neutrino see-saw provide a good effective description of observed physics below the Planck scale imposes an upper bound on the 
see-saw scale, in order to avoid sub-Planckian Landau poles. Further, this upper bound is more stringent than the bound on the scale of the dimension-five neutrino mass operator arising from partial-wave unitarity; if we attempted to set the see-saw at that scale, the neutrino Yukawa would already be non-perturbatively large at that scale and the Landau pole would be immediate. The necessity to have renormalization group running for several orders of magnitude  from the see-saw scale up to the Planck scale requires the neutrino Yukawa to be perturbatively small at the see-saw scale, and so imposes an upper bound on that scale below the bound enforced by partial-wave unitarity. The triviality bound is lower by a factor of order 2 or 3, since renormalization group running of the neutrino Yukawa is only (perturbatively) logarithmic, and it depends on the detailed field content between the see-saw scale and the Planck scale. 

%%%%%%%%%%%%%%%%%%%%%%

If we consider only the standard model field content, augmented by three generations of singlet neutrinos to implement the see-saw mechanism, and if the see-saw scale only assumes values such that the Landau pole in the see-saw sector occurs at the Planck scale or above, numerically integrating the full one loop RGEs, eq.(\ref{nsusy_rge}), leads to an upper bound on the see-saw scale of $M \approx 2\times 10^{15} \hspace{1mm} \mathrm{GeV}$ 
which is a factor 2 more stringent than that set by the requirement of partial-wave unitarity alone.

\begin{figure}[ht!]
   \newlength{\picwidtha}
   \setlength{\picwidtha}{7in}
   \begin{center}
       \resizebox{\picwidtha}{!}{\includegraphics{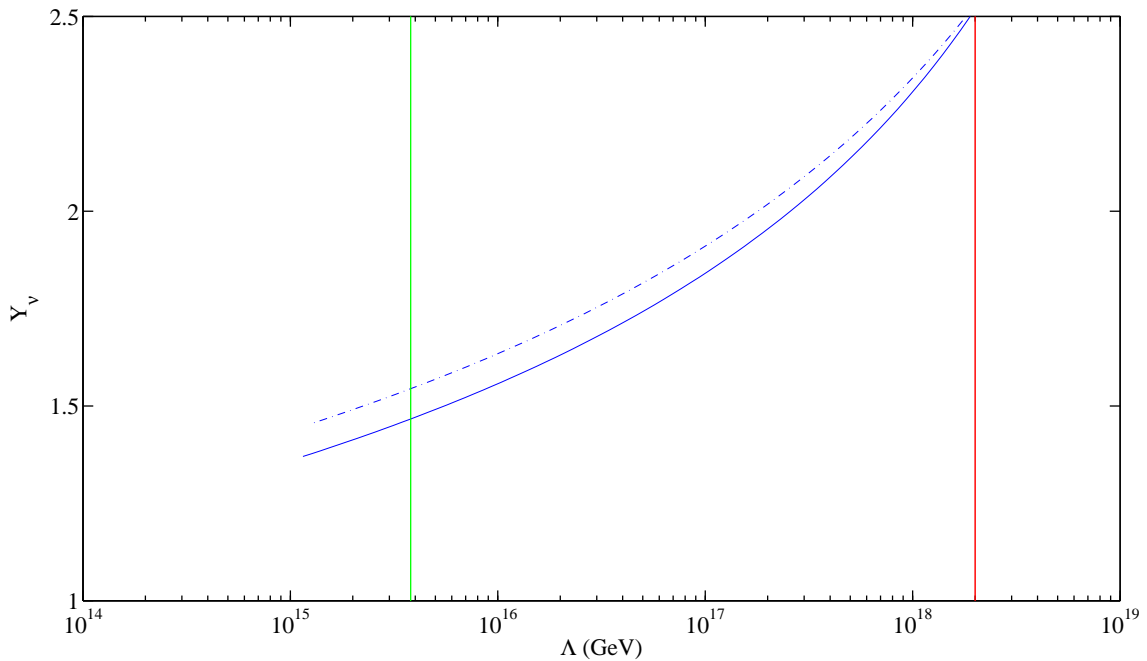}}
   \end{center}
   \caption{See-saw running of the largest eigenvalue of $Y_\nu$. 
The dashed curve corresponds to the non-supersymmetric case, the 
solid curve corresponds to the supersymmetric case (CMSSM, $\tan\beta = 10$). 
The first vertical line denotes the unitarity bound and the second vertical 
line represents the reduced Planck mass.}
   \label{Pole}
\end{figure}

%%%%%%%%%%%%%%%%%%%%%%%%%%%%%%%%%%%%%%%%%%%%%%%%%%%%%%%%%%%%%
%%%%%%%%%%%%%%%%%%%%%%%%%%%%%%%%%%%%%%%%%%%%%%%%%%%%%%%%%%%%%
As noted above, the presence of a Landau pole in the approximate solution of the renormalization group equation signals that the effective field theory has been driven into a highly non-perturbative regime, and should have been replaced by a different theory describing the effective excitations of the quanta at that scale. For a see-saw scale of $M \sim 2\times 10^{15} \hspace{1mm} \mathrm{GeV}$, the Yukawa couplings grow beyond the perturbative limit well before the scale of the Landau pole and thus the perturbative approximation no longer remains valid over the entire range up to the Planck scale. In the remainder of this paper, we make the more conservative assumption that the see-saw scale corresponds to Yukawa couplings which grow to be no larger than $\sqrt{2\pi}$ when evolved up to the Planck mass scale -- reasonably ensuring perturbativity over the (sub-Planck) range of renormalization group running. In the standard model plus see-saw case, this corresponds to a bound on the see-saw scale of $M = 1.3 \times 10^{15}\hspace{1mm} \mathrm{GeV}$ (see figure \ref{Pole}), so with these assumptions we must introduce the see-saw UV completion of the neutrino-mass effective operator at a scale at least a factor of three below the naive scale of violation of partial-wave unitarity. 
%%%%%%%%%%%%%%%%%%%%%%%%%%%%%%%%%%%%%%%%%%%%%%%%%%%%%%%%%%%%%%
%%%%%%%%%%%%%%%%%%%%%%%%%%%%%%%%%%%%%%%%%%%%%%%%%%%%%%%%%%%%%% 

%%%%%%%%%%%%%%%%%%%%%%%%%%%%%%%%%%%%%%%%%%%%%%%%%%%
%%%%%%%%%%%%%%%%%%%%%%%%%%%%%%%%%%%%%%%%%%%%%%%%%%%

Note however, that the standard model plus the see-saw not only suffers from hierarchy problems, but also does not provide a viable dark matter candidate. In the next sub-section, we will consider supersymmetric extension of this model, which provides a more realistic framework.

%%%%%%%%%%%%%%%%%%%%%%%%%%%%%%%%%%%%%%%%%%%%%%%%%%%
%%%%%%%%%%%%%%%%%%%%%%%%%%%%%%%%%%%%%%%%%%%%%%%%%%%

In much of the literature, unification has been used as a guide to determine the size and structure of the Yukawa couplings, $\Yn$, often by assuming a relationship between $\Yn$ and the up-like quark sector. There exist highly suggestive and predictive GUT scenarios, such as SO$(10)$, that naturally incorporate the see-saw mechanism in this manner (see \cite{Yan} for reviews). As a result, the largest eigenvalue of the Dirac Yukawa matrix, $\Yn$, usually corresponds to the top Yukawa coupling at the see-saw scale, up to a Clebsh-Gordon factor arising from the particulars of the GUT implementation. This model dependent determination of the Yukawa couplings generally leads to $Y_\nu \lesssim 1$ at a see-saw scale of $M \sim 10^{14}$ GeV. Using these inputs, our analysis would predict that the Landau pole develops in the deep ultraviolet, well beyond the Planck scale. Note however that grand unified theories have in general much extra field  content between the GUT scale and the Planck scale, and so the question of avoiding triviality constraints would depend on the details of the model in question and can only be determined by the analysis of the full set of RGEs for the GUT.

%%%%%%%%%%%%%%%%%%%%%%%%%%%%%%%%%%%%%%%%%%%%%%%%
%%%%%%%%%%%%%%%%%%%%%%%%%%%%%%%%%%%%%%%%%%%%%%%

%%%%%%%%%%%%%%%%%%%%%%%%%%%%%%%%%%%%%%
%%%%%%%%%%%%%%%%%%%%%%%%%%%%%%%%%%%%%%

%%%%%%%%%%%%%%%%%%%%%%%%%%%%%%%%%%%%%%%%%%%%%%%%%%%%%%%%%%%%%%%%%%%%%

\subsection{The Supersymmetric Case}

Let us now consider the framework of the supersymmetric see-saw within the CMSSM. The leptonic sector of the superpotential above the Majorana mass scale reads,
\be
\mathcal{W}={\bf Y_e}^{ij} \epsilon_{\alpha \beta} H_d^\alpha e^c_i L^\beta_j +
{\bf Y_\nu}^{ij} \epsilon_{\alpha \beta} H_u^\alpha N^c_{i} L_j^\beta +
\frac{1}{2} {\bf \mathcal{M}}^{ij} N^c_{i} N^c_{j}.
\ee
Here, $L_i$, $ i = e,\mu,\tau$, is the left handed weak doublet, $e^c_i$ is the charged lepton weak singlet, and $H_u$ and $H_d$ are the two Higgs doublets of opposite hypercharge. The SU$(2)$ indices, $\alpha$, and $\beta$, have been displayed and the anti-symmetric symbol is defined by $\epsilon_{12} =+1$. $N^c$ denotes the right-handed neutrino singlet superfield. The Yukawa matrices ${\bf Y_e}$ and ${\bf Y_\nu}$ give masses to the charged leptons and Dirac masses to the neutrinos respectively. The Majorana matrix, ${\bf \mathcal{M}}^{ij}$, gives the  right-handed neutrinos their heavy Majorana mass. In this case, the see-saw RGEs become \cite{Hisano:1995cp},
\be
\frac{d \Yn}{d t} =\frac{1}{16\pi^2}\Yn\left(3\mathrm{Tr}
\left(\Yu^\dagger \Yu \right) + \mathrm{Tr}\left(Y_\nu^\dagger \Yn \right)
+ 3 \Yn^\dagger \Yn + \Ye^\dagger \Ye - g_1^2 - 3 g_2^2 \right),
\label{susy_rge}
\ee
As the supersymmetric see-saw mechanism violates lepton number by two units, $R$-parity may be retained and hence we still have a stable LSP dark matter candidate. We will again assume a minimal extension -- only the see-saw plus the CMSSM exist below the Planck scale. As before we will assume that the Yukawa couplings saturate their perturbative limit at the Planck scale. This will have the effect of pushing the Landau pole beyond the Planck mass, while ensuring the largest possible see-saw scale. Integrating the RGEs of eq.(\ref{susy_rge}), we find that the upper bound on the see-saw scale becomes, $M = 1.2 \times 10^{15}\hspace{1mm}\mathrm{GeV}$ (see figure \ref{Pole}), again, a factor of three below the naive scale of violation of partial-wave unitarity. It should be noted that the exact determination of the see-saw scale from the one loop RGEs depends on $\tan\beta$ through the Yukawa couplings of the quarks and leptons. We take $M = 1.2 \times 10^{15}$ GeV throughout for the supersymmetric case and, provided that $5 \lesssim \tan\beta\lesssim 50$, numerically, this see-saw scale will lead to the largest eigenvalue $Y_\nu \lesssim \sqrt{2\pi}$ at $M_P$.

In addition to the superpotential, the MSSM Lagrangian contains soft supersymmetry breaking terms, namely,
\bea
-\mathcal{L}_{soft} &=& ({\bf m^2_{\tilde L}})_{ij}\tilde L^\dagger_i
\tilde L_j + ({\bf m^2_{\tilde e}})_{ij}\tilde e^*_{Ri} \tilde e_{jR} +
({\bf m^2_{\tilde \nu}})_{ij} \tilde
\nu^*_{Ri} \tilde \nu_{Rj}\nonumber \\
&& +({\bf m^2_{\tilde Q}})_{ij}\tilde Q^\dagger_i \tilde Q_j + 
({\bf m^2_{\tilde u}})_{ij}\tilde u^*_{Ri} \tilde u_{jR} + 
({\bf m^2_{\tilde d}})_{ij}\tilde d^*_{Ri} \tilde d_{jR}
\nonumber \\
&& + \tilde m_{H_d}^2 H_d^\dagger H_d + \tilde m_{H_u}^2 
H_u^\dagger H_u + (B \mu H_d H_u + \frac{1}{2} B_{\nu} 
{\bf \mathcal{M}}_{ij} \tilde \nu^*_{Ri} \tilde \nu^*_{Rj}
+\mathrm{h.c.}) \nonumber \\
&&[({\bf A_d})_{ij} H_d \tilde d^*_{Ri} \tilde Q_j + 
({\bf A_ u})_{ij} H_u \tilde u_{Ri}^* \tilde Q_j + 
({\bf A_l})_{ij}H_d \tilde e^*_{Ri}\tilde L_j
+ ({\bf A_\nu})_{ij}H_u \tilde \nu^*_{Ri}\tilde L_j
\nonumber \\
&& + \frac{1}{2}M_1 \tilde B^0_L \tilde B^0_L + \frac{1}{2} M_2
\tilde W^a_L \tilde W^a_L + \frac{1}{2} M_3 \tilde G^a \tilde G^a
+ \mathrm{h.c.}]
\label{softy}
\eea
The terms containing $\bf{m_{\tilde\nu}}^2$ and $\bf{A}_\nu$ in eq.(\ref{softy}) are only included above the see-saw scale. Model dependence enters in the choice of mechanism responsible for the soft supersymmetry breaking. An especially attractive and well motivated possibility is that supersymmetry breaking is communicated super-gravitationally from a hidden sector, in which supersymmetry is spontaneously broken, to the observable sector of the supersymmetric standard model (the MSUGRA scenario). Models with soft supersymmetry breaking masses of the form that MSUGRA would impose, and where each of the soft supersymmetry breaking scalar masses, gaugino masses and trilinear couplings are universal and flavour diagonal at the Planck scale, comprise the constrained minimal supersymmetric standard model (CMSSM) leading to the relations,
\be
({\bf m}_{\tilde f}^2)_{ij} = m_0^2 {\bf 1}
\hspace{4mm} \tilde m_{h_i}^2 = m_0^2 \hspace{4mm} 
{\bf A}_{{\bf f} ij} = am_0 {\bf Y_f},
\ee
where $m_0$ is a universal scalar mass and $a$ is a dimensionless constant. In this section, we restrict our studies to the CMSSM and set the trilinear A-term soft parameter $a=0$. For simplicity, we assume that the gaugino mass relation at the high scale reads,
\be
\frac{g_1}{M} = \frac{g_2}{M} = \frac{g_3}{M}
\ee
where $g_1$ has been appropriately normalized to the weak hypercharge. As we are not assuming a grand unified scenario, we allow the coupling constants to run past one another at the unification point $\sim 2\times 10^{16}$ GeV.

At this point it is worth examining the see-saw in more detail. After integrating out the massive right-handed neutrinos and renormalizing to the electroweak symmetry breaking scale, the Majorana mass matrix for the left-handed neutrinos becomes,
\be
{\bf m_\nu} ={\bf Y_\nu}^T {\bf \mathcal{M}}^{-1} {\bf Y_\nu} 
\langle H^0_u\rangle^2
\label{nu_mat}
\ee
where $\langle H^0_u\rangle^2 = v^2_2 = v^2 \sin^2 \beta$, and $ v= (174 \hspace{1mm} \mathrm{GeV})^2$ as set by the Fermi constant $G_F$. In the basis where ${\bf Y_e}$ and the gauge interactions are simultaneously flavor diagonal, the mass matrix eq.(\ref{nu_mat}) can be diagonalized by the PMNS matrix ${\bf U}$,
\be
{\bf U}^T {\bf m_\nu} {\bf U}= \mathrm{diag}(m_1, m_2, m_3)
\ee
The unitary PMNS matrix connects flavor eigenstates with gauge eigenstates in the lepton sector and may be parameterized as,
\be
{\bf U} = {\bf U}^\prime
\mathrm{diag} (e^{-i\phi/2}, e^{-i\phi^\prime},1)
\ee
\be
{\bf U}^\prime= \left( \begin{array}{ccc} 
c_{13}c_{12}& c_{13}s_{12}& s_{13}e^{-i\delta} \\
-c_{23}s_{12} - s_{23}s_{13}c_{12}e^{i\delta}& 
c_{23}c_{12} - s_{23}s_{13}s_{12}e^{i\delta} & s_{23}c_{13}\\
s_{23}s_{12} - c_{23}s_{13}c_{12}e^{i\delta} & 
-s_{23}c_{12} - c_{23}s_{13}s_{12}e^{i\delta} & c_{23}c_{13}
\end{array} \right).
\ee
where $\phi$ and $\phi^\prime$ are additional CP violating phases arising from the Majorana nature of the neutrino masses. We will make the simplifying assumption throughout that all of the Yukawa and mass matrices are real. In general, the low energy observables include the three mass eigenvalues and the parameters of the PMNS matrix totaling to 9 degrees of freedom. However, these low energy observables are augmented by an orthogonal matrix, $R$, \cite{Casas:2001sr} containing, in general, an additional 9 parameters. If we define,
\be
{\bf \kappa} \equiv
\frac{{\bf m_\nu}}{\langle H^0_u\rangle^2} = 
{\bf Y_\nu}^T \mathcal{M}^{-1} {\bf Y_\nu}
\ee
and use the PMNS matrix to diagonalize $\kappa$,
\be
{\bf \kappa}_d = {\bf U}^T {\bf Y_\nu}^T 
{\bf \mathcal{M}}^{-1} {\bf Y_\nu} {\bf U}.
\ee
(the $d$ subscript denotes diagonalization), and we make an arbitrary field re-definition such that ${\bf \mathcal{M}}$ appears diagonal, we arrive at
\be
{\bf 1} = \left(\sqrt{{\bf \mathcal{M}}_d^{-1}} {\bf Y_\nu} 
{\bf U} \sqrt{{\bf\kappa}_d^{-1}}\right)^T
\left(\sqrt{{\bf\mathcal{M}}_d^{-1}} {\bf Y_\nu} {\bf U} 
\sqrt{{\bf\kappa}_d^{-1}}\right).
\ee
One then identifies
\be
{\bf R} \equiv \sqrt{{\bf\mathcal{M}}_d^{-1}} {\bf Y_\nu} 
{\bf U} \sqrt{{\bf\kappa}_d^{-1}}
\ee
as an arbitrary orthogonal matrix and therefore, the most general form of ${\bf Y_\nu}$ reads \cite{Casas:2001sr}
\be
{\bf Y_\nu} = {\sqrt{{\bf \mathcal{M}}_d}} {\bf R} 
\sqrt{{\bf\kappa}_d} {\bf U}^\dagger.
\label{general}
\ee
We will see that the orthogonal matrix $R$ can have important consequences on the amount of flavor violation expected in see-saw models.

%%%%%%%%%%%%%%%%%%%%%%%%%%%%%%%%%%%%%%%%%%%%%%%%%%%%%%%%%%%%%%
%%%%%%%%%%%%%%%%%%%%%%%%%%%%%%%%%%%%%%%%%%%%%%%%%%%%%%%%%%%%%%

\section{Lepton Flavor Violation}

The standard model with massive neutrinos predicts lepton flavor violation in the charged lepton sector. However, with neutrino masses in the sub eV  range, the predicted branching ratio for $\meg$ leads to the hopelessly unobservable fraction 
%\cite{Cheng:1976uq,Cheng:1977nv,Lee:1977qz},   
\be
B(\mu \rightarrow e \gamma) = 
\frac{\Gamma(\mu \rightarrow e \gamma)}
{\Gamma(\mu \rightarrow e \nu\bar\nu)} = 
\frac{3G_F}{\sqrt{2} (32\pi)}\left(\sum_i U^*_{ei} U_{\mu i}
(m_i^2/M_W^2)\right)^2 \lesssim 10^{-40}.
\label{hopeless}
\ee
Even within our analysis of the previous section where we assumed Yukawa couplings as large as unitarity would allow, the conclusion from eq.(\ref{hopeless}) remains unaltered -- charged lepton flavor violation within the standard model with massive neutrinos is unobservable.

Lepton flavor violation proceeds through different effects in supersymmetric versions of the standard model. In the CMSSM, the soft terms can induce flavor changing neutral current processes through sfermion flavor mixing. In the scenario depicted in the previous section, flavor diagonal and universal mass spectra are imposed at the supersymmetry breaking scale, which we take as $M_P$. While universal and diagonal scalar mass terms ensure strong suppression of flavor changing neutral current effects, the CMSSM inputs at the high scale must be renormalized down to the electroweak scale. Renormalization effects will, in general, induce off-diagonal terms in the sfermion mass matrices and these terms will contribute to flavor changing neutral current processes. In the model class under study, the see-saw scale emerges below the input scale of supersymmetry breaking and thus the see-saw parameters contribute to the RGE running of the sleptons, generating off-diagonal elements in the slepton mass matrices leading to flavor violation in the charged lepton sector at testable levels \cite{Borzumati:1986qx}. 

Presently, the strongest bounds on flavor changing neutral current processes in the charged lepton sector, within the model classes we consider, come from muon decay measurements. In particular, $\meg$ provides the strongest constraint, since the rates for $\mu \rightarrow e e e$, and $\mu N \rightarrow e N$, are largely dominated by electromagnetic penguin contributions (again, in the models and parameter ranges we consider), and thus they are suppressed with respect to the rate for $\mu \rightarrow e \gamma$ by an extra factor of $\alpha$. Since at the present time the experimental limit \cite{Eidelman:2004wy} on $\mathrm{BR} (\mu \rightarrow e \gamma) \leq 1.2 \times 10^{-11} $, is of comparable strength to the limits \cite{Eidelman:2004wy} on $\mathrm{BR} (\mu\rightarrow e e e) \leq 1.0 \times 10^{-12} $, and $\mathrm{BR} (\mu N \rightarrow e N) \leq 6.1\times 10^{-13} $, the model classes that we study will be consistent with all the present lepton-flavour violation data if they satisfies the present limit on $\mathrm{BR}(\meg)$. It should be pointed out that $\mu N \rightarrow e N$ has the potential to provide the strongest constraint on charged lepton flavor violation, even with the extra factor of $\alpha$ suppression. The rate for $\mu N \rightarrow e N$ receives an enhancement from the number of nuclei in the atom and studies have indicated the experimental feasibility \cite{Popp:2004ds} of obtaining a limit of $\mathrm{BR} (\mu N \rightarrow e N) \sim 10^{-18}$, which, in the case of penguin domination, would correspond to $\mathrm{BR} (\meg) \sim 10^{-15}$.

In the leading log approximation to the RGE running of the soft supersymmetry breaking terms within the see-saw sector, and assuming sfermion dominance, the branching ratio for $\meg$ becomes:
\bea
\mathrm{BR}(\mu \rightarrow e \gamma) &\simeq& \frac{\alpha^3}
{G_F^2} \frac{(m^2_{{\bf L}})_{12}^2}{m_s^8} \tan^2 \beta \nonumber \\
&\simeq& \frac{\alpha^3}{G_F^2 m_s^8} \left| \frac{1}{8\pi^2}
(3+ a^2)m_0^2\ln\frac{M_{X}}{M_R}\right|^2 \left|\left({\bf Y_\nu}^\dagger
{\bf Y_\nu}\right)_{12}\right|^2 \tan^2 \beta
\label{BR}
\eea
where $m_s$ denotes the mass of a typical sfermion.

Notice that the branching ratio eq.(\ref{BR}) depends on see-saw parameters through $\left|\left({\bf Y_\nu}^\dagger {\bf Y_\nu}\right)_{12}\right|^2$. The structure of $\Yn$ therefore plays a central role in determining the amount of LFV expected. Fortunately, the parameterization of eq.(\ref{general}) allows us to easily explore two important limiting cases -- a strong mass hierarchy among the right-handed neutrinos, or a degenerate mass relations between the right-handed neutrinos. 
 
%%%%%%%%%%%%%%%%%%%%%%%%%%%%%%%%%%%%%%%%%%%%%%%%%%%%%%%%%%
%%%%%%%%%%%%%%%%%%%%%%%%%%%%%%%%%%%%%%%%%%%%%%%%%%%%%%%%%%

\section{LFV in the CMSSM}

If we assume a strong mass hierarchy among the $\nu_R$s, the Majorana mass matrix in the diagonal basis reads, $\mathcal{M} \approx \mathrm{diag}(0,0,M_R)$, implying the expression 
(eq.(\ref{general}))
\be
({\bf Y_\nu})_{ij} = \sqrt{\mathcal{M}_3}
\delta_{i3}{\bf R}_{3l} (\sqrt{{\bf \kappa}_d})_l {\bf U}^\dagger_{lj}.
\label{Hi}
\ee
The appearance of ${\bf R}_{3l}$ in eq(\ref{Hi}) defines an important feature in this limiting case and the parameter can be phrased through the dependence on a single angle, $\theta_1$, originating from the orthogonal matrix ${\bf R}$ \cite{Casas:2001sr}. In particular, the element of interest that determines the amount of LFV expected from eq.(\ref{BR}) reads 
\bea
F(\theta_1) \equiv |({\bf Y_\nu}{\bf Y_\nu}^\dagger)_{12}|^2 &\approx& 
\frac{Y_0^2}{\kappa_2 \sin^2\theta_1 + 
\kappa_3\cos^2\theta_1}\left|(\sqrt{\kappa_2} U_{12}^* \sin\theta_1 + 
\sqrt{\kappa_3} U_{13}^*\cos\theta_1)\right. \nonumber \\
&& \left. \times (\sqrt{\kappa_2} U_{22} \sin\theta_1 + 
\sqrt{\kappa_3} U_{23}\cos\theta_1)\right|^2
\label{angle_eq}
\eea
\begin{figure}[ht!]
   \newlength{\picwidthg}
   \setlength{\picwidthg}{7in}
   \begin{center}
       \resizebox{\picwidthg}{!}{\includegraphics{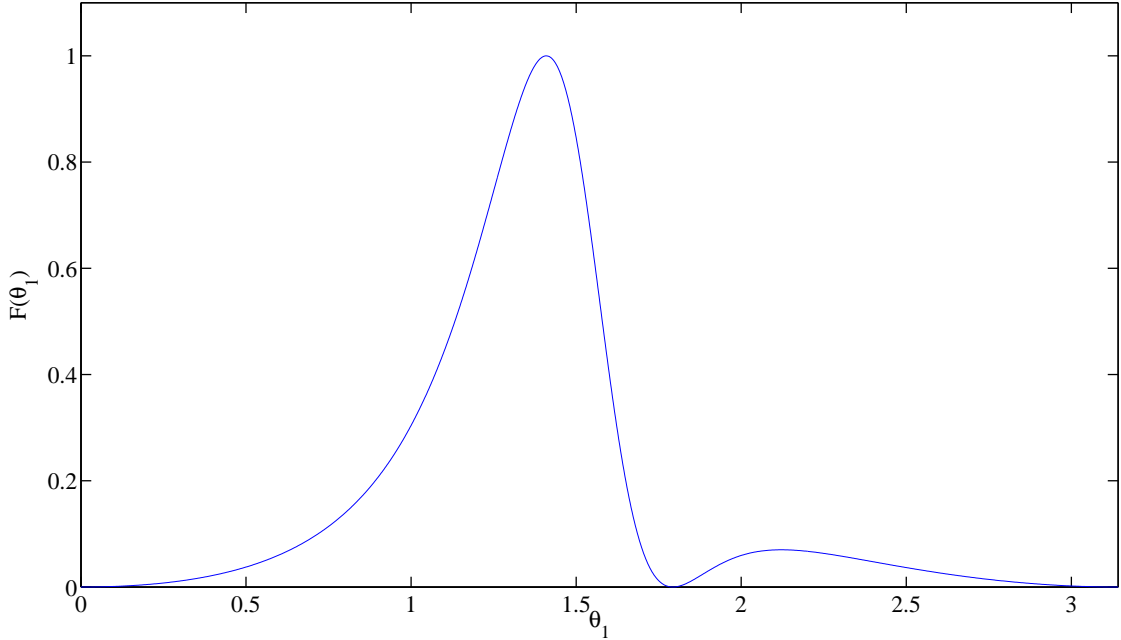}}
   \end{center}
   \caption{$F(\theta_1)$ in arbitrary units normalized to the 
global maximum. Approximately $30\%$ of $\theta_1$ corresponds to the 
global maximum within a factor of two.}
   \label{peak}
\end{figure}
where $Y_0^2$ represents the largest eigenvalue of $({\bf Y_\nu}{\bf Y_\nu}^\dagger)$, determined by the see-saw scale. The parameters are defined as, $\kappa_2= \sqrt{\Delta m_{12}^2}$ and $\kappa_3 = \sqrt{\Delta m_{23}^2}$. We assume a strong hierarchy among the left-handed neutrinos throughout and we take the LMA solution for the PMNS matrix. The function, $F(\theta_1)$, takes on a global maximum near $\theta_1 = 1.4$ and approximately $30\%$ of the range of $\theta_1$ corresponds to the the global maximum within an order of magnitude (see figure \ref{peak}). Therefore, according to eq.(\ref{BR}), approximately $30\%$ of the parameter range of $\theta_1$ will correspond to a branching ratio for $\meg$ within one order of magnitude of the maximum predicted value.

\begin{figure}[ht!]
   \newlength{\picwidthb}
   \setlength{\picwidthb}{7in}
   \begin{center}
       \resizebox{\picwidthb}{!}{\includegraphics{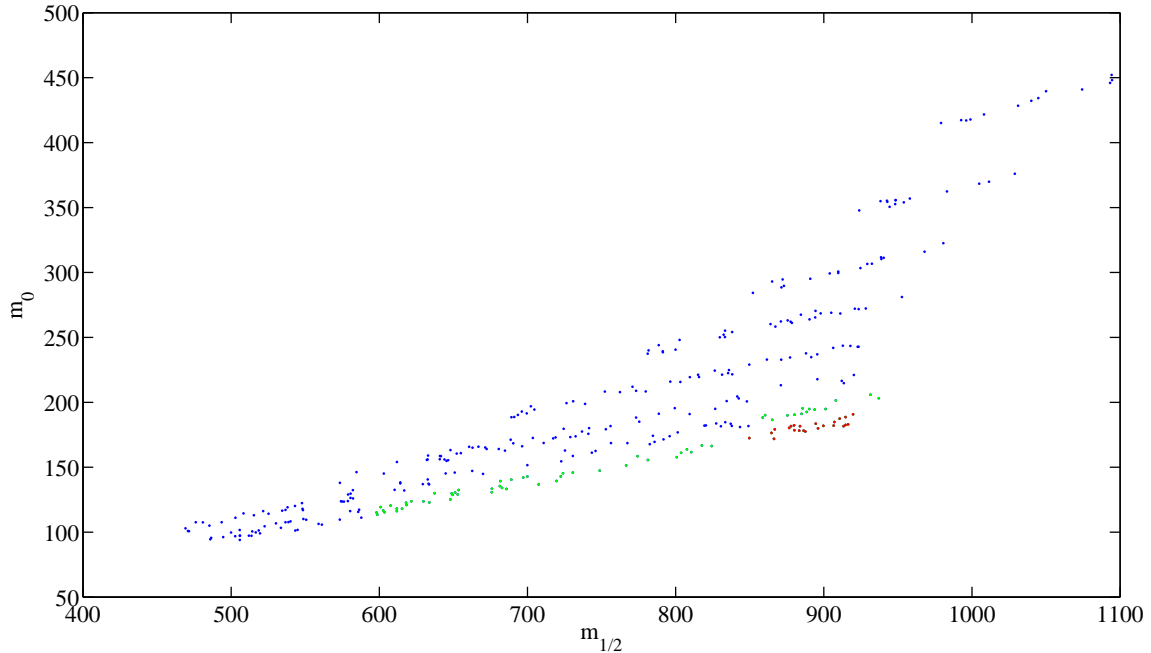}}
   \end{center}
   \caption{Approximate CMSSM region left after LFV constraints imposed, 
$\tan\beta=5$--$40$. Percent of $\theta_1$ parameter space allowed: 
Red $25\%$ -- $20 \%$, green $20\%$ -- $10\%$, 
blue $10 \%$ -- $5 \%$; $\mu >0$}
   \label{percent}
\end{figure}

If we consider the allowed CMSSM parameter space, consistent with all laboratory and cosmological constraints, we find that most of angle $\theta_1$ becomes eliminated for a see-saw scale of $\Lambda=1.2 \times 10^{15}\hspace{1mm}\mathrm{GeV}$. The angle $\theta_1$ becomes increasingly restricted to regions near the minimum of eq.(\ref{angle_eq}). As figure \ref{percent} demonstrates, less than $20 \%$ of $\theta_1$ is allowed in the most optimistic cases and in most of the CMSSM parameter range, less than $10 \%$ of $\theta_1$ is allowed. In no permitted region is more than $25\%$ of $\theta_1$ allowed. This places severe constraints on this framework.

If we assume a degenerate spectrum of right-handed neutrinos the relevant term to $\mathrm{BR}(\meg)$ becomes
\be
{\bf Y_\nu}^\dagger {\bf Y_\nu} = {\bf \mathcal{M}} {\bf U} 
{\bf \kappa}_d {\bf U}^\dagger,
\ee
and the dependence on the angle $\theta_1$ is lost. In this case, the prediction for the branching ratio is fixed by the low energy parameters of the PMNS matrix, the light neutrino masses, and the see-saw scale. We find that the entire range of CMSSM parameter space is eliminated in the degenerate $\nu_R$ case.

%%%%%%%%%%%%%%%%%%%%%%%%%%%%%%%%%%%%%%%%%%%%%%%%%%%%%%%%%%%%
%%%%%%%%%%%%%%%%%%%%%%%%%%%%%%%%%%%%%%%%%%%%%%%%%%%%%%%%%%%%

\section{LFV in Split Supersymmetry}

Recently, much interest has focused on a particular realization of supersymmetry known as split supersymmetry \cite{Arkani-Hamed:2004fb,Giudice:2004tc,Arkani-Hamed:2004yi}. In these models the gaugino and higgsino mass spectrum remains light (TeV range) and the A-terms remain small, protected from large masses by $R$-symmetry, while the sfermions acquire large (from multiple TeV to GUT scale) masses at the supersymmetry breaking scale. While this scenario no longer uses supersymmetry to solve the gauge hierarchy problem, this model class retains $R$-parity and hence a dark matter candidate, and can also preserve grand unification, at least at one loop \cite{Arkani-Hamed:2004yi}. Moreover, the large sfermion masses strongly suppress flavor changing neutral currents and CP violating effects, removing observational tension in the CMSSM. While these models do have aesthetically unpleasant features, well motivated studies of wide classes of constructions demonstrate that a split supersymmetic spectrum can arise
\cite{Antoniadis:2004dt,Kors:2004hz,Dudas:2005vv,Dudas:2005pr}, and split supersymmetry provides unique predictions, e.g. \cite{Arkani-Hamed:2004yi,Masiero:2004ft,Giudice:2005rz}.

In the CMSSM, ($m_s \sim \mathrm{TeV}$), with a see-saw scale suggested by grand unification ($\Lambda \sim 10^{14}\hspace{1mm} \mathrm{GeV}$), 
the predicted branching ratio for $\meg$ is close to the current experimental bound in most of the allowed parameter range. Since naively the branching ratio falls off like $1/m_s^4$, a moderate increase in the scalar mass spectrum would suppress the branching ratio significantly, placing the prediction below the reach of upcoming experimental searches. However, as the previous section demonstrated, the branching ratio rises sharply for large Yukawa couplings (or equivalently large see-saw scales). In addition to the enhancement of the branching ratio from large Yukawa couplings, the branching ratio $1/m_s^4$ dependence becomes invalid near scalar-gaugino mass degeneracy. In this case, the estimate given in eq.(\ref{BR}) no longer holds. Numerically, the the gaugino contributions in this region of parameter space suppresses the branching ratio through cancellation effects. The $1/m_s^4$ dependence does not set in until the scalar masses become sufficiently large ($m_{0} \sim 10 m_{1/2}$). 

If large Yukawa couplings are assumed, scalar mass terms require large values to drive the branching ratio down to acceptable levels. We determine the scale of scalar masses that can be probed with LFV in this set-up. In what follows, we assume a split supersymmetry scenario with large universal scalar masses and light universal gaugino and higgsino masses imposed at the Planck scale. As split supersymmetry requires $R$-symmetry to protect the gaugino and higgssino masses, the framework naturally suggests D-term suprsymmetry breaking (F-term breaking will lead to the spontaneous violation of $R$-symmetry). Since D-term breaking provides a soft supersymmetry breaking mass spectrum proportional to gauge charges, a natural expectation is the loss of universality at the input scale. Nevertheless, it is possible to construct models \cite{Dudas:2005vv,Dudas:2005pr} of split supersymmetry with universal scalar masses and, for simplicity, we will assume a universal scalar mass spectrum at the supersymmetry breaking scale, $M_P$. Furthermore, we will enforce a fine-tuning in the Higgs sector to accommodate one light Higgs boson and the correct $Z$-boson mass. In all cases we take $\mu >0$ as the sign of the $\mu$-parameter does not play a significant role in the LFV rate. Numerically, we integrate the MSSM RGEs from the Planck scale down to the weak scale, integrating out each $\nu_R$ at its associated mass and integrating out the sfermions at the split 
supersymmetry scale. The off-diagonal elements in the slepton mass matrices stop running at the split supersymmetry scale and the size of the elements at this scale determine the amount of LFV predicted. In addition, we ensure that only one Higgs acquires a negative mass squared, ensuring radiative electroweak symmetry breaking while preserving electromagnetic charge and colour. 

\begin{figure}[ht!]
\newlength{\picwidthc}
\setlength{\picwidthc}{4.2in}
 \begin{center}
\subfigure[][]{\resizebox{\picwidthc}{!}{\includegraphics{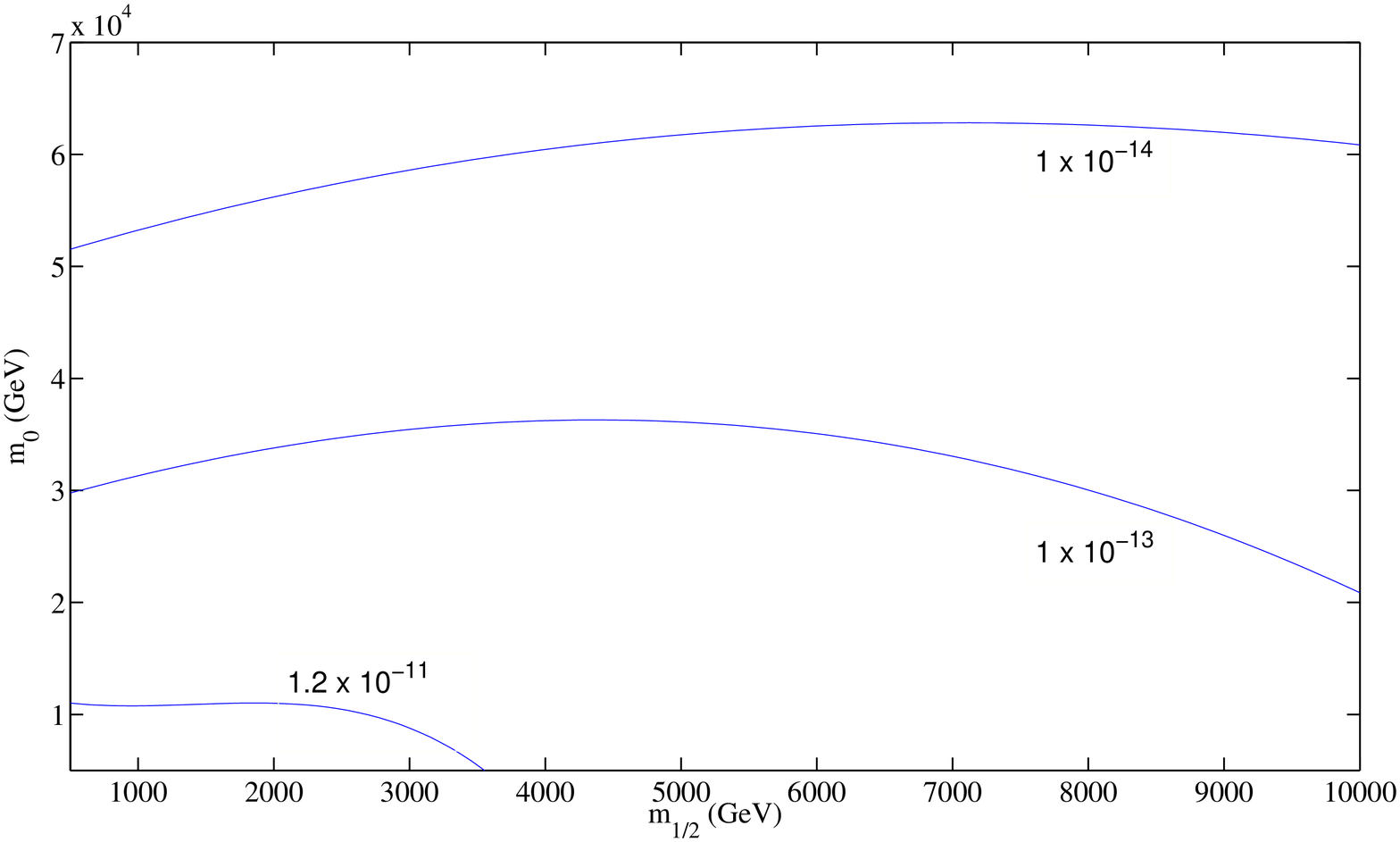}}}
 \subfigure[][]{\resizebox{\picwidthc}{!}{\includegraphics{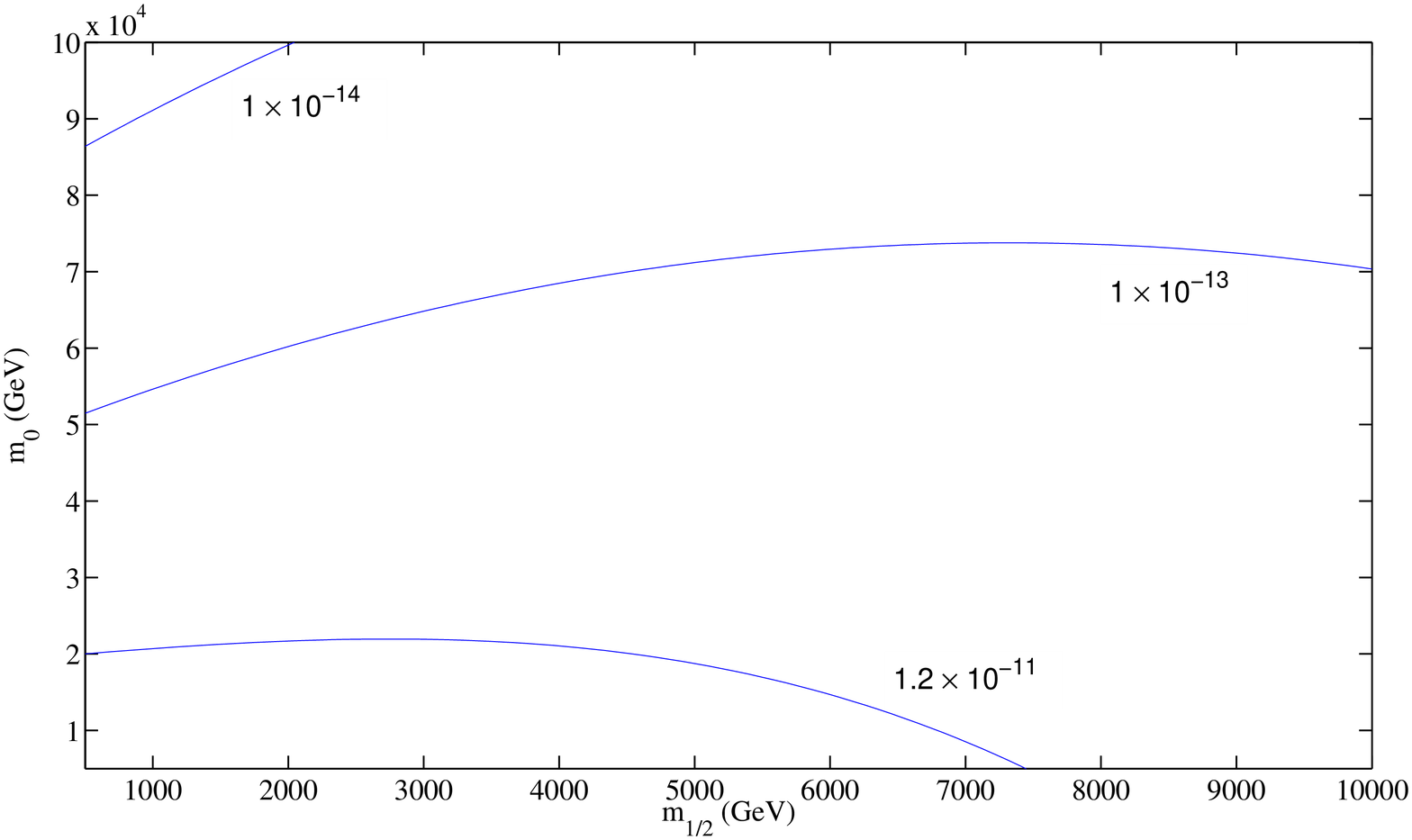}}}
 \subfigure[][]{\resizebox{\picwidthc}{!}{\includegraphics{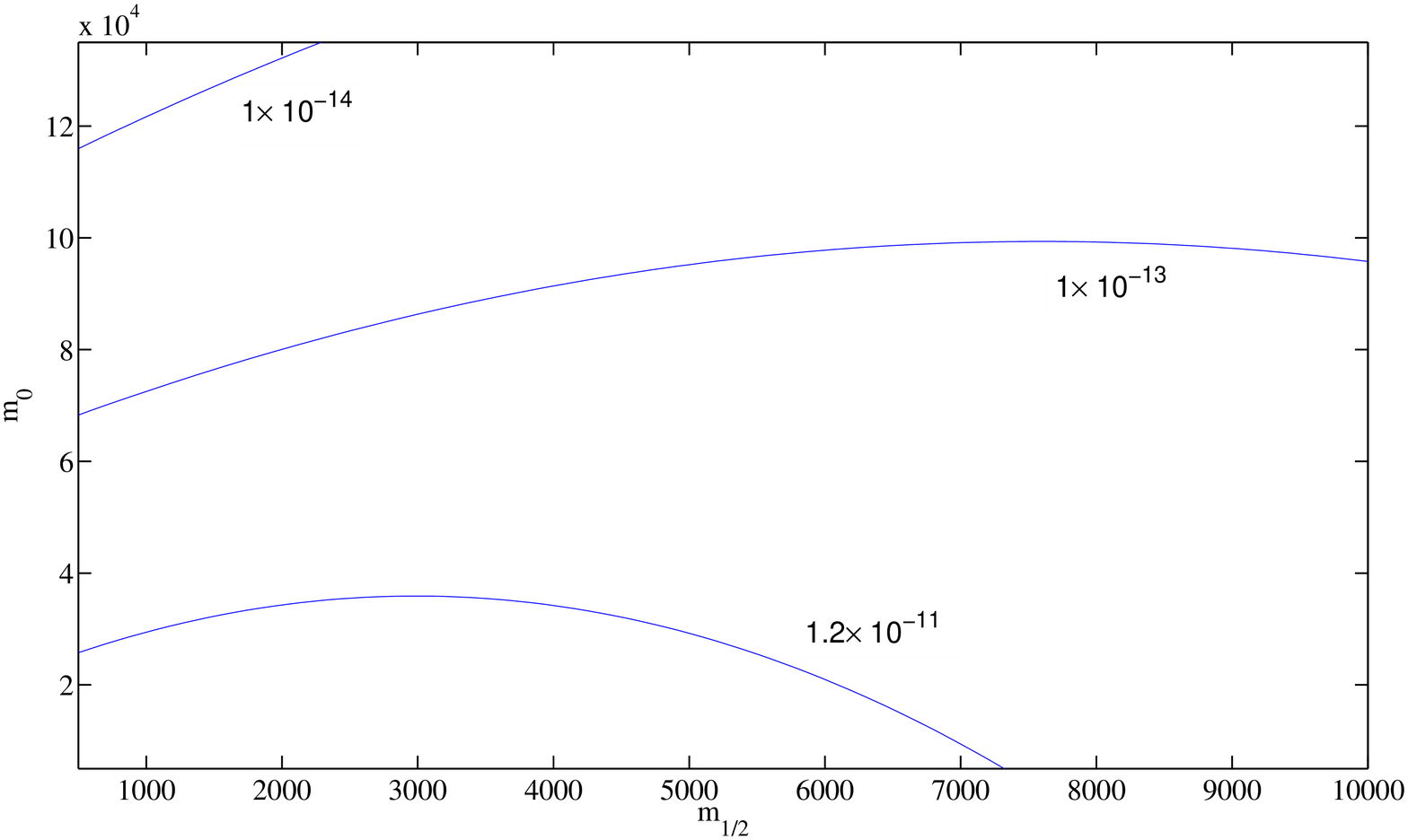}}}
 %LAST GRAPH HAS LABELING ERROR, NOW FIXED
 \end{center}
\caption{Contours of constant branching ratio, $m_{1/2} =1000$ GeV, 
$\mu >0$, $\nu_R$ hierarchical. a) -- c) include $\tan\beta=5,20,40$ 
with $\theta_1=1.4$}
\label{tanb1}
\end{figure}

\begin{figure}[ht!]
\newlength{\picwidthh}
\setlength{\picwidthh}{4.2in}
 \begin{center}
 \subfigure[][]{\resizebox{\picwidthh}{!}{\includegraphics{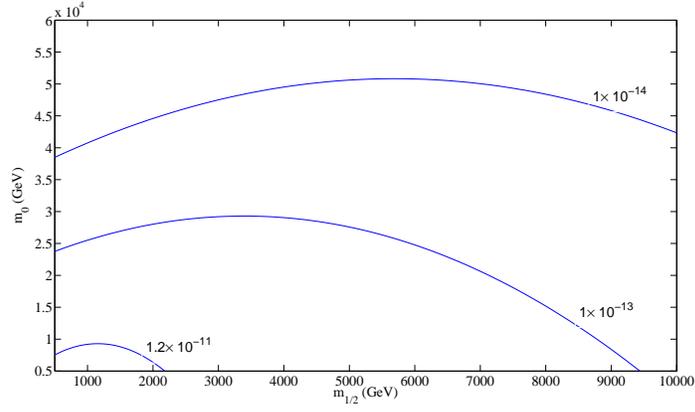}}}
  \subfigure[][]{\resizebox{\picwidthh}{!}{\includegraphics{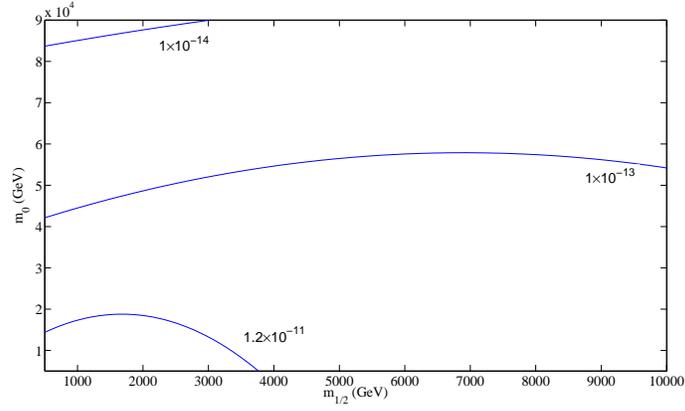}}}
   \subfigure[][]{\resizebox{\picwidthh}{!}{\includegraphics{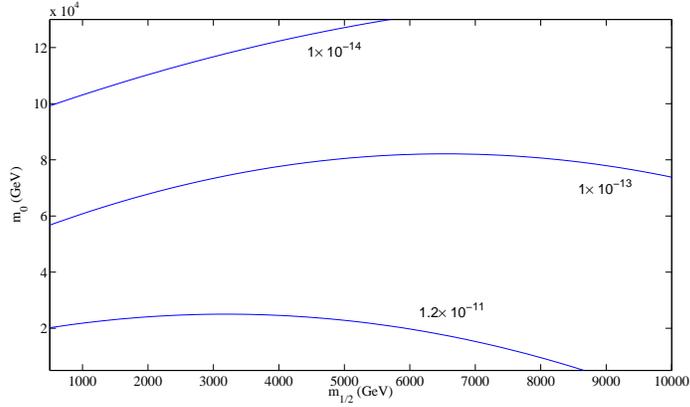}}}
 \end{center}
\caption{Contours of constant branching ratio, $m_{1/2} =1000$ GeV, 
$\mu >0$, $\nu_R$ degenerate. a) -- c) include $\tan\beta=5,20,40$.}
\label{tanb2}
\end{figure}

\begin{figure}[ht!]
\newlength{\picwidthd}
\setlength{\picwidthd}{4.2in}
 \begin{center}
\subfigure[][]{\resizebox{\picwidthd}{!}{\includegraphics{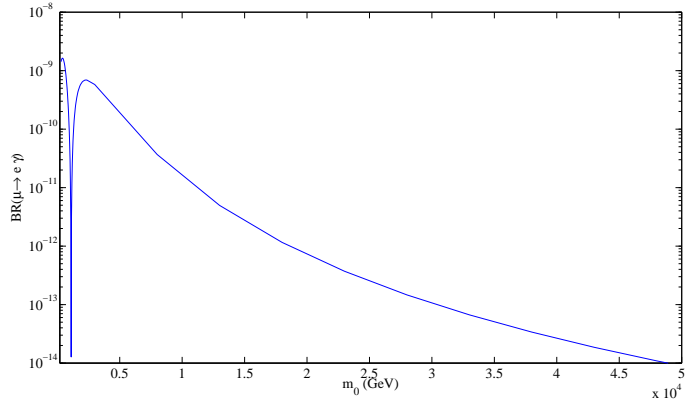}}}
 \subfigure[][]{\resizebox{\picwidthd}{!}{\includegraphics{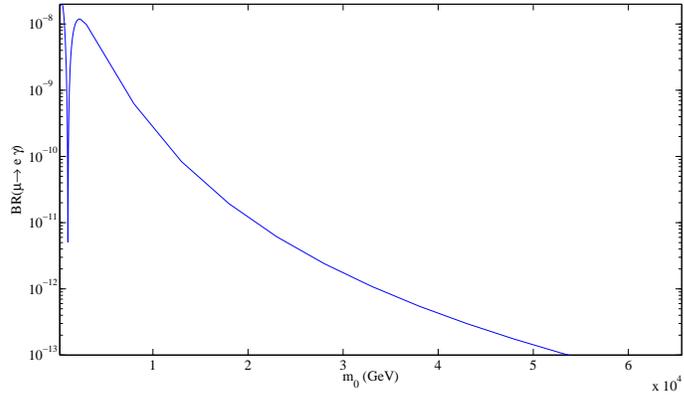}}}
 \subfigure[][]{\resizebox{\picwidthd}{!}{\includegraphics{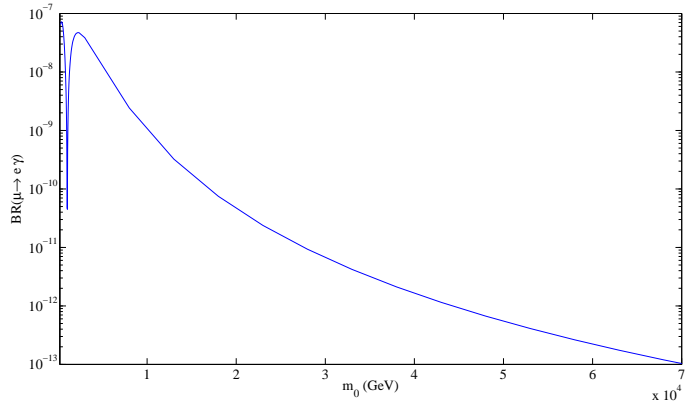}}}
 \end{center}
\caption{Branching ratio vs $m_0$, $m_{1/2} = 1000$ GeV, $\mu >0$, 
$nu_R$ hierarchical. a) -- c) correspond to $\tan\beta =5,20,40$ 
respectively with $\theta_1 =1.4$}
\label{BRM1}
\end{figure}

\begin{figure}[ht!]
\newlength{\picwidthk}
\setlength{\picwidthk}{4.2in}
 \begin{center}
 \subfigure[][]{\resizebox{\picwidthk}{!}{\includegraphics{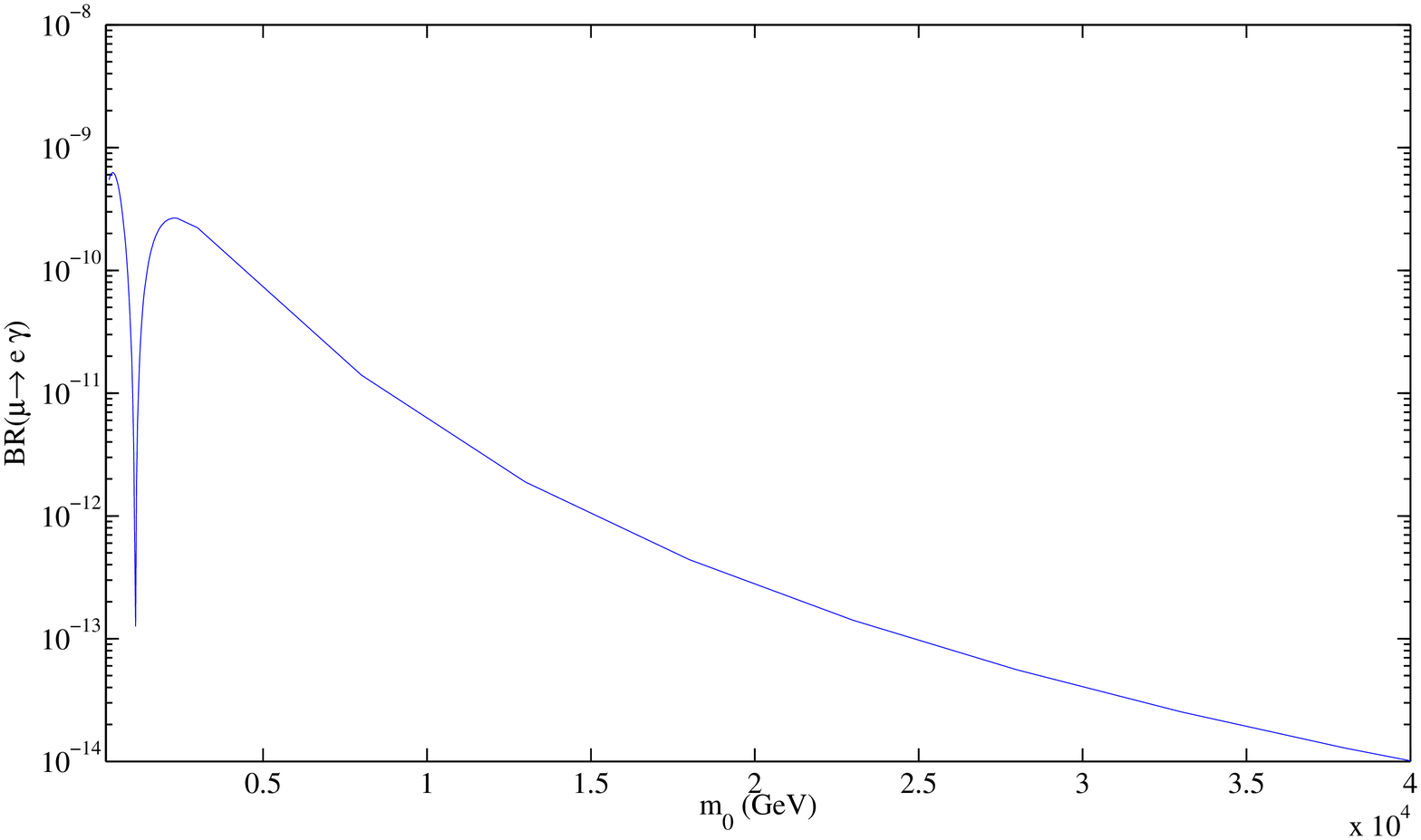}}}
  \subfigure[][]{\resizebox{\picwidthk}{!}{\includegraphics{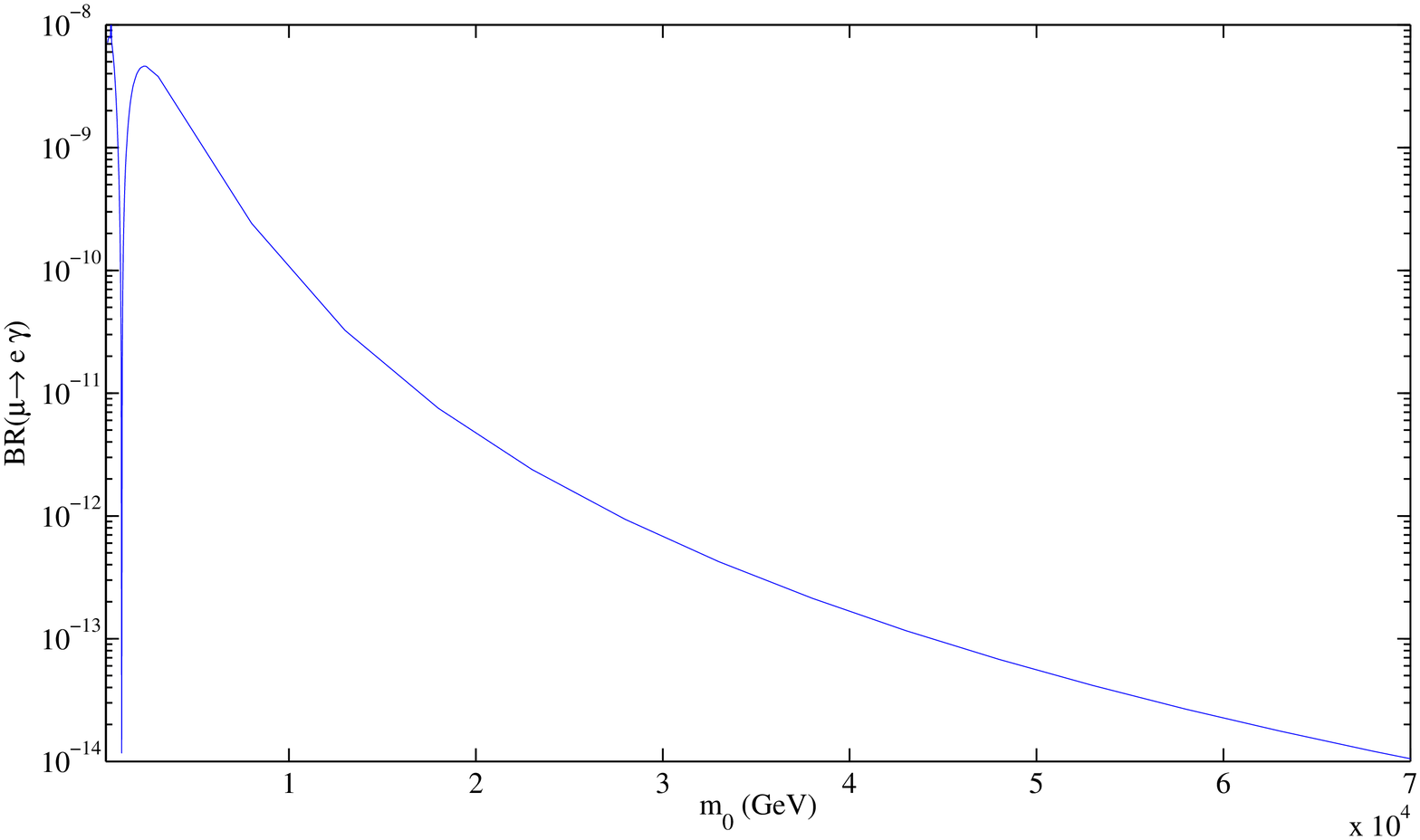}}}
   \subfigure[][]{\resizebox{\picwidthk}{!}{\includegraphics{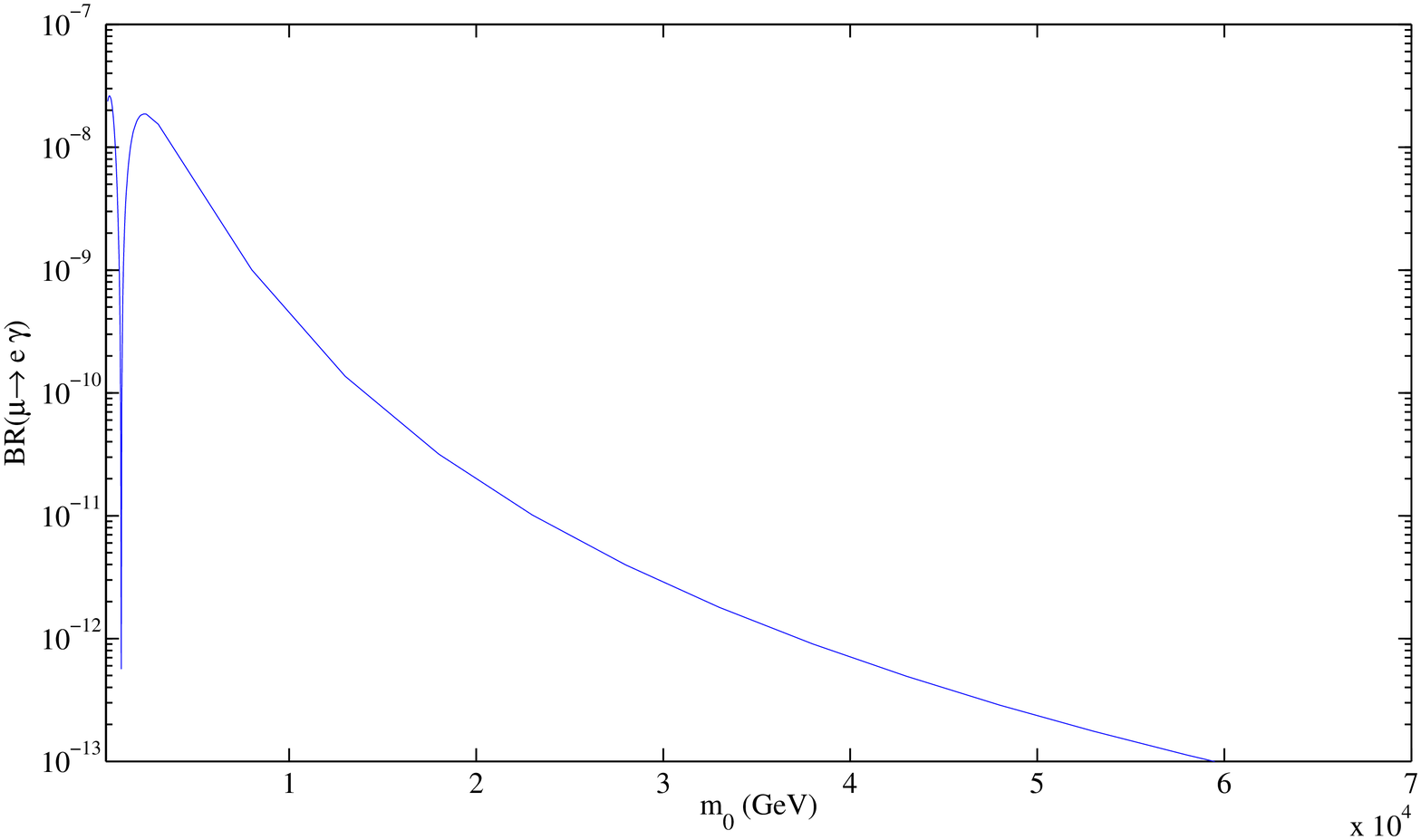}}}
 \end{center}
\caption{Branching ratio vs $m_0$, $m_{1/2} = 1000$ GeV, $\mu >0$, 
$\nu_R$ degenerate. a) -- c) correspond to $\tan\beta =5,20,40$ respectively.}
\label{BRM2}
\end{figure}

We find that LFV can probe scales as high as $\sim 100$ TeV in this scenario. In figure \ref{tanb1}, and figure \ref{tanb2} we display contours of constant branching ratio in universal scalar mass and universal gaugino mass space. Figure \ref{tanb1} displays the hierarchical case for $\tan\beta=5,20,40$ with $\theta_1=1.4$. With $\theta_1$ chosen at this value, we explore a region of parameter space that corresponds to the maximum LFV. It should be noted that $0.8<\theta_1<1.5$ corresponds to LFV levels within one order of magnitude the maximum value. In figure \ref{tanb2}, we show the analogous case with degenerate $\nu_R$ where the is no dependence on $\theta_1$. In figure \ref{BRM1}, and figure \ref{BRM2}, we display the branching ratio as a function of the universal scalar mass, assuming $m_{1/2} =1000 \hspace{1mm} \mathrm{GeV}$ and $\mu>0$. Subfigures a) through c) in figure \ref{BRM1} display the hierarchical $\nu_R$ case with $\theta_1 = 1.4$ and subfigures a) through c) in figure \ref{BRM2} show the degenerate case. Note that the graphs do not come below the $1\times 10^{-13}$ level 
until the universal scalar mass approaches $50$ TeV in most cases. Upcoming searches of the MEG experiment \cite{Mori:2002sg} expect to explore BR($\meg$) at the $1 \times 10^{-13}$ level which, as indicated from figures \ref{tanb1}--\ref{BRM2}, corresponds to scalar masses $30$ -- $80$ TeV range. The $100$ TeV range is of particular interest as bottom up considerations have shown that PeV scale ($10^3$ TeV) split supersymmetry presents a highly suggestive phenomenology \cite{Wells:2004di}. In these models, much of the parameter space yields scalar masses in the $100$ TeV range. We should note split supersymmetry does not predict an absolute masses scale for supersymmetry breaking and therefore constructions with scalar masses $m_0 \gg 10^5$ GeV are in principle possible. With scalar masses this large, even with large see-saw Yukawa couplings, $\meg$ would not be expected with an observable rate. However, intermediate split supersymmetry scales, $m_0 = 10^6$ -- $10 ^9$ GeV, are disfavoured \cite{Ibarra:2005vb} (provided that universality is relaxed) as much of the parameter space yields tachyonic squark masses (spontaneously breaking colour) and cosmological considerations suggest that $m_0 \lesssim 10^{12}$ -- $10^{13}$ GeV \cite{Arkani-Hamed:2004fb,Hewett:2004nw}. Thus the most favoured split supersymmetry scenarios appear to include a relatively moderately split spectrum with $m_0 \sim 10^3$ -- $10^6$ GeV, which has a potential sensitivity to $\meg$, or a strongly split spectrum with scalar masses $\sim 10^{10}$ -- $10^{13}$ GeV. It should be stressed that we have not made any assumptions other than a see-saw scale below, but near, the triviality bound, universal supersymmetry breaking at $M_P$, and that we may ignore other possible new physics that may exist below the Planck scale. 
If split supersymmetry with $m_0 \sim 100 \hspace{1mm} \mathrm{TeV}$ is realized in nature and $\meg$ is discovered at levels of the MEG searches, one of the implications includes a see-saw scale near the triviality bound, {\it i.e.} large Yukawa couplings in the see-saw sector. We emphasize 
that we conservatively assume a see-saw scale that retains perturbativity over the entire range of RGE running.

%%%%%%%%%%%%%%%%%%%%%%%%%%%%%%%%%%%%%%%%%%%%%%%%%%%%%%%%%%%%%%%%%
%%%%%%%%%%%%%%%%%%%%%%%%%%%%%%%%%%%%%%%%%%%%%%%%%%%%%%%%%%%%%%%%%

\section{Conclusions}

We have used the requirement of the absence of a Landau pole in the see-saw sector below the Planck scale (``triviality" bounds) to determine bounds on the see-saw scales in both the supersymmetric and non-supersymmetric see-saw extensions of the standard model. By demanding that the see-saw Yukawa couplings remain perturbative up to the Planck scale, we have established an upper bound on the see-saw scale of $\Lambda = 1.3 \times 10^{15}$ GeV in the non-supersymmetric version of the see-saw, and $1.2\times10^{15}$ GeV in the supersymmetric case. See-saw scales near these bounds lead to large Yukawa couplings over the domain of renormalization group running up to $M_P$, which in the supersymmetric case, with gravity-mediated supersymmetry breaking, leads to strong constraints from lepton flavor violation.

In the CMSSM, most of the allowed parameter range (figure \ref{percent}) is eliminated for a see-saw scale of $1.2\times10^{15}$ GeV with hierarchical $\nu_R$s and completely eliminated in the degenerate $\nu_R$ case. This places strong constraints on the theory. If the LHC establishes the CMSSM, the low level of charged-lepton flavor violation implies a see-saw scale much below the bounds given by triviality considerations for most 
of the see-saw parameter range. However, in supersymmetric models with large scalar masses, such as split supersymmetry, lepton flavor violation can be used as a probe of see-saw scales near the triviality limit. If $\meg$ is observed at levels that can be probed in the MEG experiment, $\mathrm{BR}(\meg) \sim 10^{-13}$, and if a moderately split supersymmetric spectrum ($\sim$ PeV) is realized with supergravity mediated supersymmetry breaking scalar masses and with see-saw generated neutrino masses, the implications would include a high see-saw scale near its triviality bound.  

%%%%%%%%%%%%%%%%%%%%%%%%%%%%%%%%%%%%%%%%%%%%%%%%%%%%%%%%%%%%%%%%%
%%%%%%%%%%%%%%%%%%%%%%%%%%%%%%%%%%%%%%%%%%%%%%%%%%%%%%%%%%%%%%%%%

\section{Acknowledgments}
We would like to thank Antonio Masiero and Sudhir Vempati for useful discussions. We would also like to acknowledge the support of the Natural Sciences and Engineering Research Council of Canada.

%%%%%%%%%%%%%%%%%%%%%%%%%%%%%%%%%%%%%%%%%%%%%%%%%%%%%%%%%%%%%%%%%%
%%%%%%%%%%%%%%%%%%%%%%%%%%%%%%%%%%%%%%%%%%%%%%%%%%%%%%%%%%%%%%%%%%
 
\bibliography{ref}

%\section{References}

%%%%%%%%%%%%%%%%%%%%%%%%%%%%%%%%%%%%%%%%%%%%%%%%%%%%%%%%%%%%%%%

\end{document}